\documentclass[12pt]{article}
\usepackage{amsmath,amssymb,latexsym,psfig}
\usepackage{color}
\usepackage[all]{xy}

\def\sqr#1#2{{\vcenter{\vbox{\hrule height.#2pt
            \hbox{\vrule width.#2pt height#1pt \kern#1pt
                  \vrule width.#2pt}\hrule height.#2pt}}}}

\def\square
 {\mathop{\mathchoice{\sqr{12}{15}}{\sqr{9}{12}}
{\sqr{6.3}{9}}{\sqr{4.5}{9}}}}
\begin{document}

\hfill hep-th/0502027

\vspace{0.5in}
 
\begin{center}

{\large\bf Notes on gauging noneffective group actions}

\vspace{0.25in}
                                                                               
Tony Pantev$^1$ and Eric Sharpe$^{2}$ \\
$^1$ Department of Mathematics \\
University of Pennsylvania \\
David Rittenhouse Lab.\\
209 South 33rd Street \\
Philadelphia, PA  19104-6395\\
$^2$ Departments of Physics, Mathematics \\
University of Utah \\
Salt Lake City, UT  84112 \\
{\tt tpantev@math.upenn.edu},
{\tt ersharpe@math.utah.edu} \\

$\,$

\end{center}

In this paper we study sigma models in which a
noneffective group 
action has been gauged.  Such gauged sigma
models turn out to be different from gauged sigma models in which an
effectively-acting group is gauged, because of nonperturbative effects
on the worldsheet.  We concentrate on finite noneffectively-acting
groups, though we also outline how analogous phenomena also happen in
nonfinite noneffectively-acting groups.  We find that understanding
deformations along twisted sector moduli in these theories leads one
to new presentations of CFT's, defined by fields valued in roots of unity.

\begin{flushleft}
January 2005
\end{flushleft}
                                                                               
\newpage
                                                                               
\tableofcontents
                                                                               
\newpage

\section{Introduction}
                     
In this paper we shall collect some results on the physics of gauged
sigma models in which a noneffectively-acting group has been gauged.
By ``noneffectively-acting,'' we mean that some nontrivial elements of
the group act trivially, {\it i.e.} $g \cdot x = x$ for all $x$, for
some $g$ other than the identity.  Such trivially-acting elements form
a normal subgroup, call it $K$, of the gauge group $G$, and so as
$G/K$ is a group, the reader might suspect that a $G$-gauging would be
physically equivalent to a $G/K$ gauging.  However, that is not the
case, as we shall see in numerous examples.  Gauging $G$ is a distinct
physical operation from gauging $G/K$, because of 
nonperturbative effects.

In this paper, we shall concentrate on understanding finite
noneffectively-acting groups.  We will briefly outline how analogous
phenomena also happen when gauging nonfinite noneffectively-acting
groups, but a more extensive discussion will appear in \cite{tonyme,glsm}.
We will discuss massless spectrum computations in such orbifolds,
which have features that make them a bit more subtle than ordinary
orbifolds by effectively-acting groups.  We shall also discuss various
other technical issues in noneffective orbifolds, such as possible
D-branes, and, to a limited extent, mirror symmetry.  (Mirror symmetry
in noneffective gaugings will be discussed much more extensively in
\cite{glsm}.)  Curiously, we shall see that twist fields associated to
trivially-acting group elements are often equivalent to fields valued
in roots of unity, a fact which will play an important part in the
sequels \cite{tonyme,glsm}, where we will rederive the same
description from completely independent lines of reasoning.

Part of the purpose of this paper is to lay part of the physical
groundwork for the upcoming papers \cite{tonyme,glsm}, which will
describe what it means to compactify a string on a general Calabi-Yau
stack.  In a nutshell, under some mild conditions,
every stack\footnote{When we speak of stacks,
we will always assume the stacks are smooth,
complex, algebraic, Deligne-Mumford stacks.  This means, for example,
that all $G$-actions will be assumed to have finite stabilizers, throughout
this paper as well as \cite{tonyme,glsm}.  We will explain the mild conditions
alluded to above in \cite{tonyme}.} has a
presentation of the form $[X/G]$ for some manifold $X$ and some group
$G$ with an action on $X$, where $G$ need not be finite and need not
act effectively.  To such a presentation $[X/G]$, one associates a
$G$-gauged sigma model on $X$.  Thus, studying string
compactifications on stacks boils down to studying gauged sigma
models.  The first important point is that a stack can have many
presentations of the form $[X/G]$, which can define very different
gauged sigma models.  For example, if $G$ is finite in one
presentation and nonfinite in another, then in the first presentation,
the gauged sigma model is a CFT, whereas typically in the second
presentation, the gauged sigma model will not be a CFT.  Thus, stacks
cannot classify gauged sigma models; rather, the most one can hope
for is that universality classes\footnote{These are
equivalence classes of gauged sigma models, where two such models are
declared equivalent if they are related by worldsheet RG flow.} of
gauged sigma models are classified by stacks.  Such a claim cannot be
checked directly, but numerous indirect tests are possible, as we
shall describe in \cite{tonyme,glsm}.  In those papers we also resolve
various obstacles to the consistency of this claim,
perhaps most importantly the mismatch between physical CFT
deformations and mathematical deformations of stacks.  Finally, in
order to understand both the resolution of the puzzles posed by
deformation theory, as well as mirror symmetry, one is led to study 
new presentations of abstract CFT's defined by fields valued in roots of
unity.

In this paper, we shall concentrate on the physics of gauging
noneffective (and primarily finite) group actions, though we shall
occasionally use the language of stacks to help make contact with the
sequels \cite{tonyme,glsm}.  Among other things, in this paper we
will see that twist fields associated to trivially-acting group elements
are equivalent to fields valued in roots of unity; we will also recover
such fields from completely independent lines of reasoning in
\cite{tonyme,glsm}.

We begin in section~\ref{basicexs} with a discussion of several
examples of gauged finite noneffectively-acting groups.  We explicitly
compute that gauging a noneffectively-acting group is distinct from
gauging an effectively-acting one, and also check
that these noneffective gaugings are consistent -- the theories are
modular-invariant, for example.  Mathematically, the types of stacks
associated with noneffective gaugings are known as `gerbes,' and since
we will be using the language of stacks in \cite{tonyme,glsm}, we
relate our examples to that mathematical language.

In section~\ref{nonmincharge} we briefly outline analogous phenomena
in gauging noneffective nondiscrete groups.  A much more extensive
discussion of such gaugings, together with numerous examples and
computations, will appear in \cite{tonyme,glsm}; for the purposes of
this paper, we merely point out the existence of analogous phenomena
there.

In section~\ref{spectra} we discuss massless spectrum computations in
orbifolds by noneffectively-acting finite groups.  We find that the
massless spectrum has the same general form as for finite
effectively-acting groups, {\it i.e.} one twisted sector for each
conjugacy class, even if the elements of that class act trivially.
However, the reasoning behind this result is a bit subtle, and since
we have not seen a detailed explanation of massless spectra in
noneffective orbifolds in the physics literature previously, we spend
a great deal of time discussing potentially confusing issues.  One of
the more important issues is understanding the physical infinitesimal
deformations dictated by the results of the massless spectrum
computation.  In typical cases, the noneffective orbifold has more
(unobstructed) moduli fields than its effectively-acting
counterpart, but the only ones that have a clear geometric
meaning are that subset in the effectively-acting theory.  This
issue can be restated more formally as a mismatch between the number
of physical moduli of gauged sigma models and the number of
mathematical moduli of the stack.  As the physical moduli are, by
definition, part of the massless spectrum, it is very important to
understand this issue to properly understand the massless spectrum.
 
In section~\ref{defthy} we discuss this issue and outline how resolving it
leads to new presentations of CFT's.  In \cite{tonyme} such deformation theory
issues will play a much more important role, and will be discussed much
more extensively.

One prerequisite for the deformation theory discussion is to rewrite
twist fields for trivially-acting group elements in a
different-looking fashion.  In section~\ref{trivgerbesection} we
discuss how such twist fields are equivalent to fields valued in roots
of unity, which gives us a very algebraic description of many of the
twisted sector fields.  Such a description is impossible for a twist
field associated to a nontrivially-acting group element, but
trivially-acting group elements are special in this regard.  We also
take the oppotunity in this section to compare an orbifold of a space
$X$ by a trivially-acting ${\bf Z}_k$ to the CFT of a sigma model on
$k$ disjoint copies of $X$.  The two theories are distinct, but do
share certain features.

In section~\ref{Dbranenoneff} we discuss D-branes in noneffective orbifolds.
Even if a group element acts trivially on the space,
it can still act nontrivially on the Chan-Paton factors, as this is
consistent with the Cardy condition.

In section~\ref{mirrors} we briefly discuss mirror symmetry in the special
case of noneffective orbifolds in which the entire orbifold group acts
trivially.  We will discuss mirror symmetry for gauged sigma models much more
extensively in \cite{glsm}.

The noneffective gaugings we discuss in this paper all have
multiple dimension zero operators in their spectru, which signals
a failure of cluster decomposition.  Such a failure is not fatal
in two-dimensional conformal field theories, however,
as for example a sigma model on a disjoint union of spaces also
has multiple dimension zero operators, and so also fails cluster
decomposition.  These issues are not unrelated;
in the followup work \cite{clusterdecomp} we shall argue that
the conformal field theories obtained by these noneffective gaugings
are equivalent to conformal field theories describing disjoint unions
of spaces.

\section{Examples of global quotients by finite
noneffectively-acting groups}   \label{basicexs}

Consider an orbifold of a space $X$ by a group $G$,
containing elements that act trivially.  As
mentioned in the introduction, those elements form a normal subgroup
of $G$, call it $K$.  Now, the reader might at first glance suspect
that gauging $G$ would be physically equivalent to gauging $G/K$, but
we shall see in examples in this section that this
is not the case, by computing one-loop partition functions (and,
incidentally, checking modular invariance).  Gauging a
noneffectively-acting group is not the same as gauging an
effectively-acting group.

In \cite{tonyme} we will give such orbifolds an alternative
interpretation, as examples of sigma models\footnote{To be precise,
a sigma model is defined on a presentation of a stack, not precisely on
the stack itself.  In the discussion in the paragraph above, each orbifold
is canonically associated with a particular presentation, and sigma models
are defined on those presentations.} on Calabi-Yau gerbes.
A gerbe is, in essence, a local orbifold by a trivially-acting group,
and can be presented as global quotients by larger
noneffectively-acting groups.
A ``Calabi-Yau gerbe'' is a gerbe that can be presented as
a quotient of a Calabi-Yau by a $G$-action that preserves
the holomorphic top-form.
If $K$ lies in the center of $G$, we say the gerbe is `banded.'
If $K$ does not lie in the center of $G$, we say the gerbe is
`non-banded.'  We shall see that this banded versus non-banded distinction
is reflected in the one-loop partition functions.

\subsubsection{First example:  trivially-acting ${\bf Z}_2$ center}
\label{ex:d4elliptic}

We shall begin by considering a family of examples in which 
the trivially-acting subgroup is $K = {\bf Z}_2$,
and with the 
property that $K$ lies in the center of $G$.
To be specific, take $G = D_4$, the eight-element dihedral group,
which is 
a nonabelian group that can be described
as a\footnote{Nontrivial central extensions of
${\bf Z}_2 \times {\bf Z}_2$ by ${\bf Z}_2$
are not unique. For example, the quaternions define another 
eight-element nonabelian group, which is such
a central extension.  The quaternions and  $D_4$ are not isomorphic,
as can be checked by {\it e.g.} comparing the orders of their elements.}
nontrivial central extension of ${\bf Z}_2 \times {\bf Z}_2$
by ${\bf Z}_2$:
\begin{equation}  \label{Hext}
1 \: \longrightarrow \: {\bf Z}_2 \: \longrightarrow \: D_4 \:
\longrightarrow \:
{\bf Z}_2 \times {\bf Z}_2 \: \longrightarrow \: 1
\end{equation}
We can alternately
describe $D_4$ as upper-triangular
$3 \times 3$ matrices with $1$'s on the diagonal and the strictly
upper-triangular elements in ${\bf F}_2$.
Define
\begin{displaymath}
z \: = \: \left[ \begin{array}{ccc}
                 1 & 0 & 1 \\
                 0 & 1 & 0 \\
                 0 & 0 & 1 \end{array} \right], \: \:
a \: = \: \left[ \begin{array}{ccc}
                 1 & 1 & 0 \\
                 0 & 1 & 0 \\
                 0 & 0 & 1  \end{array} \right], \: \:
b \: = \: \left[ \begin{array}{ccc}
                 1 & 1 & 1 \\
                 0 & 1 & 1 \\
                 0 & 0 & 1  \end{array} \right]
\end{displaymath}
It is straightforward to check that everything commutes with $z$
(in fact, the image of ${\bf Z}_2$ in $D_4$ is $\{ I, z \}$ where
$I$ denotes the $3 \times 3$ identity matrix), and that
\begin{displaymath}
z^2 \: = \: a^2 \: = I, \: \: b^2 \: = \: z, \: \:
ba \: = \: abz
\end{displaymath}
The eight group elements are
\begin{displaymath}
I, \: z, \: a, \: b, \: ab, \: az, \: bz, \: ba
\end{displaymath}
and $H$ has five conjugacy classes, given by
\begin{displaymath}
\{ I \}, \: \{ z \}, \: \{ a, az \}, \: \{ b, bz \}, \:
\{ab, ba \}
\end{displaymath}

Let $D_4$ act on a manifold $X$ by first projecting to ${\bf Z}_2 \times
{\bf Z}_2$, and then letting the ${\bf Z}_2 \times {\bf Z}_2$ act,
so that the ${\bf Z}_2$ center acts trivially.  Gauging the $D_4$ action
means that we must sum over principal $D_4$ bundles on the worldsheet,
so that, for example, the one-loop partition function of a $D_4$ gauged
sigma model has the same form as if the $D_4$ were acting effectively:
\begin{displaymath}
Z(D_4) \: = \: \frac{1}{|D_4|} \sum_{g,h \in D_4, gh=hg} Z_{g,h}
\end{displaymath}

How does this string orbifold $[ X/D_4 ]$ compare to the string orbifold
defined by the stack $[X/ ({\bf Z}_2 \times {\bf Z}_2)]$?
It is straightforward to check that, for each twisted sector
of $[X/ ({\bf Z}_2 \times {\bf Z}_2)]$ there are $| {\bf Z}_2 |^2 = 4$
twisted sectors
in $[X/D_4]$ with the same boundary conditions on the fields,
{\it except} for the\footnote{Each one-loop twisted sector is defined by
an equivalence class of principal $G$-bundles over an elliptic curve,
or, equivalently, a pair of commuting group elements.  We denote such
one-loop twisted sectors by a square with the commuting group elements on the
sides, corresponding to a representation of the elliptic curve as a square
with sides identified, and marking how the commuting group elements appear.
} 
\begin{displaymath}
{\scriptstyle a} \square_b \, , \: \: \:
{\scriptstyle a} \square_{ab} \, , \: \: \:
{\scriptstyle b} \square_{ab}
\end{displaymath}
twisted sectors
of $[X/ ({\bf Z}_2 \times {\bf Z}_2)]$.
Since there is no way to lift those pairs of group elements to
commuting pairs of group elements in $D_4$, there are no corresponding
one-loop twisted sectors.  The one-loop partition functions of the two theories
are related as
\begin{displaymath}
Z(D_4) \: = \: \frac{ | {\bf Z}_2 \times {\bf Z}_2 | }{ |D_4|} |{\bf Z}_2|^2
 \left[ Z({\bf Z}_2 \times {\bf Z}_2 ) \: - \:
\left( \mbox{some twisted sectors} \right) \right]
\end{displaymath}
Moreover, it is easy to check in examples that the omitted one-loop
twisted sectors are nonzero in general.
Thus, this physical theory is
distinct from the orbifold $[X/({\bf Z}_2 \times {\bf Z}_2)]$,
with a manifestly different partition function.
Gauging the noneffectively-acting
$D_4$ is {\it not} the same as gauging the effectively-acting
$D_4/{\bf Z}_2 = {\bf Z}_2 \times {\bf Z}_2$.
                                                                                
Omitting twisted sectors from a string orbifold partition function
runs the risk of destroying modular invariance.  After all,
in a one-loop partition function,
\begin{displaymath}
\left[ \begin{array}{cc}
       m & n \\
       p & q \end{array} \right] \: \in \: SL(2, {\bf Z})
\end{displaymath}
sends the
\begin{displaymath}
{\scriptstyle g} \square_h
\end{displaymath}
twisted sector to the
\begin{displaymath}
{\scriptstyle g^m h^n} \square_{g^p h^q}
\end{displaymath}
twisted
sector.
                                                                                
In the case of the $[X / ({\bf Z}_2 \times {\bf Z}_2)]$ orbifold, however,
it is nonetheless possible to omit some of the twisted sectors without
destroying modular invariance, and such a truncation is precisely what we
have obtained from our noneffective orbifold.  The omitted twisted sectors
precisely fill an $SL(2,{\bf Z})$ orbit.
None of the remaining
twisted sectors can be mapped into the omitted twisted sectors
by the $SL(2,{\bf Z})$ action, so our $[X/D_4]$ orbifold is modular-invariant.                                                                                
In hindsight, the fact that modular invariance is not broken should
not surprise us.  If we think of the orbifold as a $[X/D_4]$ quotient,
and keep track of the non-effectively-acting part of the group,
then the partition function is manifestly modular invariant -- we sum
over all commuting pairs of elements of $D_4$.  It is only when we try
to think of the orbifold in terms of some operation on a ${\bf Z}_2 \times
{\bf Z}_2$ orbifold that modular invariance becomes more obscure.
More generally, whenever one has an $[X/G]$ quotient for $G$ finite
but not necessarily effectively-acting, the one-loop partition function
will contain copies of some of the twisted sectors in an $[X/H]$
orbifold for some effectively-acting $H$,
and the multiplicities between different one-loop
twisted sectors might vary, but the resulting partition function will
always be modular invariant.

So far we have not specified the manifold $X$,
but examples are easy to construct.  One well-known example of a space
with a ${\bf Z}_2 \times {\bf Z}_2$ action is $T^6$.
One can then define a $D_4$ action on $T^6$ by first projecting to
${\bf Z}_2 \times {\bf Z}_2$, and then letting the
${\bf Z}_2 \times {\bf Z}_2$ act.

\begin{figure}
\centerline{\psfig{file=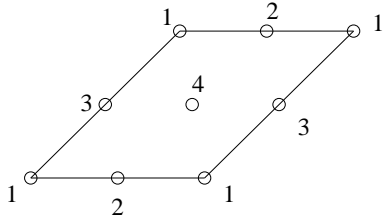,width=2in}}
\caption{\label{2tr} The four two-torsion points on an elliptic curve.}
\end{figure}

We can also construct examples in which the ${\bf Z}_2 \times {\bf Z}_2$
acts freely.  For example, ${\bf Z}_2 \times {\bf Z}_2$ has a free
action on $T^2$, defined by translations by the
2-torsion points on $T^2$, \cite[section II.1]{silvtate}
${\bf Z}_2 \times {\bf Z}_2$.  These points are sketched in
figure~\ref{2tr}.

We have previously mentioned that such quotients by noneffectively-acting
groups are examples of gerbes.  Since we will be using that language
more extensively in the follow-up works \cite{tonyme,glsm},
let us explore what that means in the
present example. 
Since the ${\bf Z}_2$ subgroup acts trivially, the quotient is a
${\bf Z}_2$-gerbe over $T^2/ ({\bf Z}_2 \times {\bf Z}_2)$,
{\it i.e.} a local orbifold by a trivially-acting ${\bf Z}_2$,
and it can be shown that it is a nontrivial ${\bf Z}_2$ gerbe.
Also, since the ${\bf Z}_2$ lies in the center of $D_4$, this is
a banded ${\bf Z}_2$ gerbe, and since the group action preserves
the holomorphic top-form, it is a Calabi-Yau banded ${\bf Z}_2$ gerbe.
                                                                         
It is instructive to note that this is a nontrivial
gerbe. 
Let $X$ be any Calabi-Yau, with an action of ${\bf Z}_2 \times
{\bf Z}_2$.  Then, in particular, $X$ is a principal
${\bf Z}_2 \times {\bf Z}_2$ bundle over the stack
$[ X / ( {\bf Z}_2 \times {\bf Z}_2 ) ]$, and as such,
is classified by an element
\begin{displaymath}
\xi \: \in \: H^1\left( [ X/ ( {\bf Z}_2 \times {\bf Z}_2 ) ],
{\bf Z}_2 \times {\bf Z}_2 \right)
\end{displaymath}
The short exact sequence~(\ref{Hext}) induces a long exact sequence
containing a map
\begin{displaymath}
H^1\left( [ X/ ( {\bf Z}_2 \times {\bf Z}_2 ) ],
{\bf Z}_2 \times {\bf Z}_2 \right)
\: \longrightarrow \:
H^2\left( [ X/ ( {\bf Z}_2 \times {\bf Z}_2 ) ],
{\bf Z}_2 \right)
\end{displaymath}
and the image of $\xi$ under this map is the characteristic
class of the gerbe we are currently interested in.
This characteristic class of the gerbe will be trivial
if and only if $\xi$ is in the image of the map
\begin{equation} \label{eq:mapD4}
H^1\left( [ X/ ( {\bf Z}_2 \times {\bf Z}_2 ) ],
D_4 \right)
\: \longrightarrow \:
H^1\left( [ X/ ( {\bf Z}_2 \times {\bf Z}_2 ) ],
{\bf Z}_2 \times {\bf Z}_2 \right)
\end{equation}
{\it i.e.} if and only if the principal ${\bf Z}_2 \times {\bf Z}_2$
bundle $X$ lifts to a principal $D_4$ bundle over $[X/\left( {\bf Z}_2 \times
{\bf Z}_2\right)]$.  In the present case,
${\bf Z}_2 \times {\bf Z}_2$ acts freely and
$Y:= [X/\left({\bf Z}_2 \times {\bf Z}_2\right)]$ is
a smooth elliptic 
curve. 
But it is easy to see that there are no such
principal $D_4$ bundles.  Every such $D_{4}$ bundle
  corresponds to a 
representation $\rho : \pi_{1}(Y) \to D_{4}$, such that
$\operatorname{im}(\rho) \subset D_{4}$ surjects onto ${\bf Z}_2
\times {\bf Z}_2$. However $D_{4}$ does not contain any abelian
subgroups that surject onto  $D_{4}$.  Thus
$\xi$ cannot be in the image of the map \eqref{eq:mapD4}, and so the
characteristic 
class must be nontrivial.

\subsubsection{Another set of ${\bf Z}_2$ gerbes}
                                                                               
Another example of a ${\bf Z}_2$ gerbe can be built
from an orbifold $[X/D_8]$ where the ${\bf Z}_2$ center of
$D_8$ acts trivially.
In general, the group $D_{n}$ is generated by two elements,
call them $a$, $b$, subject to the relations
\begin{displaymath}
a^2=1, \: \: b^{n} = 1, \: \: aba \: = \: b^{-1}
\end{displaymath}
The center of $D_{2n}$ is ${\bf Z}_2$, generated by $b^n$,
and $D_{2n}/{\bf Z}_2 = D_n$, {\it i.e.},
\begin{displaymath}
1 \: \longrightarrow \: {\bf Z}_2 \: \longrightarrow \:
D_{2n} \: \longrightarrow \: D_n \: \longrightarrow \: 1
\end{displaymath}
is exact, and describes $D_{2n}$ as a (nontrivial) central extension of $D_n$.
In an orbifold $[X/D_8]$ where the center of $D_8$ acts trivially,
so that only $D_4$ acts effectively on $X$,
we find that the resulting theory looks much like a $D_4$ orbifold,
except that some one-loop twisted sectors are omitted,
but the remaining theory is still modular invariant.
The remaining one-loop twisted sectors follow the pattern that
for any $g \in D_4$ along one side, the allowed group elements on
the other side are
generated by $g$.  Since the ${\bf Z}_2$ lies in the center,
and so each remaining one-loop twisted sector appears in the $D_8$ orbifold
with multiplicity $| {\bf Z}_2 |^2$, the one-loop partition function has
the form
\begin{displaymath}
Z(D_8) \: = \: \frac{| D_4 | }{| D_8 |} | {\bf Z}_2 |^2 \left[
Z(D_4) \: - \: \left( \mbox{some one-loop sectors} \right) \right]
\end{displaymath}
the same general form as in the previous example.

\subsubsection{${\bf Z}_2$ gerbe over a dihedral orbifold}

Another example of a gerbe over a space can be obtained as follows.
Let $DD_n$ denote the binary dihedral group, generated (as a subgroup
of $SL(2,{\bf C})$) by the matrices
\begin{displaymath}
a \: = \: \left[ \begin{array}{cc}
                 \xi & 0 \\
                 0 & \xi^{-1} \end{array} \right], \: \:
b \: = \: \left[ \begin{array}{cc}
                 0 & 1 \\
                -1 & 0 \end{array} \right]
\end{displaymath}
for $\xi$ an $n$th root of unity, obeying relations including
$b^2 = -I$ and $aba = b$.  For simplicity,
we shall assume $n=3$.
This group projects onto a representation of the dihedral
group in $SL(3,{\bf C})$, generated by the matrices
\begin{displaymath}
a' \: = \: \left[ \begin{array}{ccc}
                 \xi & 0 & 0 \\
                 0 & 1 & 0 \\
                 0 & 0 & \xi^{-1} \end{array} \right], \: \:
b' \: = \: \left[ \begin{array}{ccc}
                  0 & 0 & 1 \\
                  0 & -1 & 0 \\
                  1 & 0 & 0 \end{array} \right]
\end{displaymath}
obeying relations including $b'^2 = + I$ and $a' b' a' = b'$.
If we denote the first group by $DD$ and the second by $D$,
then as the first projects onto the second with kernel ${\bf Z}_2$,
we have a short exact sequence
\begin{displaymath}
1 \: \longrightarrow \: {\bf Z}_2 \: \longrightarrow \:
DD \: \longrightarrow \: D \: \longrightarrow \: 1
\end{displaymath}
that we can use to define a ${\bf Z}_2$ gerbe over an orbifold
$[X/D]$.  For example, consider $X = E \times A \times E$,
where $E$ and $A$ are both elliptic curves, and $E$ has a complex
multiplication by a  cube root of unity.
Let $D$ act on $E \times A \times E$ in the obvious way, via its
description in terms of $SL(3,{\bf C})$ matrices. 

It is not hard to check that this gerbe is not
trivial as well.  Indeed, a trivialization of this gerbe is the same
thing as a principal ${\bf Z}_{2}$-bundle on $E \times A \times E$
which is equipped with an action of $DD$ that lifts the natural action
of $D$ on $E \times A \times E$. Every such ${\bf Z}_{2}$-bundle
corresponds to a character $\pi_{1}(E\times A\times E) \to {\bf
Z}_{2}$ which is invariant under the natural action of $D$ on
$\pi_{1}(E\times A\times E)$. Using the identifications $\pi_{1}(E)
\cong \pi_{1}(A) \cong {\bf Z}^{2}$ one checks immediately that $a'$
and $b'$ act on $\pi_{1}(E\times A\times E) = {\bf Z}^{2}\oplus {\bf
  Z}^{2}\oplus {\bf Z}^{2}$ via the block matrices
\[
\begin{pmatrix} {\mathbb X} & & \\ & {\mathbb I} & \\ & & {\mathbb X} 
\end{pmatrix} \qquad \text{and} \qquad 
\begin{pmatrix}
& & {\mathbb I} \\ & - {\mathbb I} & \\ {\mathbb I} & & , 
\end{pmatrix}
\]
respectively. Here ${\mathbb I}$ is the $2\times 2$ identity matrix
and ${\mathbb X}$ is the $2\times 2$ matrix
\[
{\mathbb X} = \begin{pmatrix} 0 & -1 \\ 1 & -1 \end{pmatrix}.
\]
Using this explicit form for the $D$ action on $\pi_{1}(E\times
  A\times E)$, one checks directly that the only $D$-invariant ${\bf
  Z}_{2}$-valued character of $\pi_{1}(E\times A\times E)$ is the
  trivial one. Thus $[X/DD]$ will be trivial only if $D\times {\bf
  Z}_{2}$ admits a surjective homomorphism onto $DD$, which is clearly
  impossible. One can also check that in $[X/DD]$,
  at one-loop one gets exactly $| {\bf Z}_2 |^2$ copies of each
  one-loop twisted sector of $[X/D]$, so the resulting one-loop
  partition function is manifestly modular-invariant, and the closed
  string massless spectrum of $[X/DD]$ is the same as that of $[X/D]$.

\subsubsection{A non-banded gerbe}

So far we have only discussed banded gerbes, {\it i.e.} the
trivially-acting part of the group has been central.
Let us next consider a more general example.
Let ${\bf H}$ denote the eight-element group of
quaternions, {\it i.e.}
\begin{displaymath}
{\bf H} \: = \: \{ \pm 1, \pm i, \pm j, \pm k \}
\end{displaymath}
Consider a nontrivial ${\bf Z}_4$ gerbe over
the orbifold $[X/{\bf Z}_2]$ constructed by using the fact
that ${\bf H}$ can be expressed as
\begin{displaymath}
1 \: \longrightarrow \: \langle i \rangle \: \longrightarrow \: {\bf H}
\: \longrightarrow \: {\bf Z}_2 \: \longrightarrow \: 1
\end{displaymath}
where $\langle i \rangle$ denotes the cyclic subgroup of order four generated
by $i \in {\bf H}$.
The subgroup $\langle i \rangle$ is not in the center of ${\bf H}$,
hence this extension is not central, and so the gerbe
$[X/ {\bf H}]$ (in which ${\bf H}$ acts by first projecting to
${\bf Z}_2$ and then using the given ${\bf Z}_2$ action)
is not a ${\bf Z}_4$-banded gerbe, but merely a ${\bf Z}_4$ gerbe.
Nontrivial Calabi-Yau gerbes of this form can be constructed by {\it e.g.}
taking $X$ to be a Calabi-Yau with $\pi_1$ containing a
${\bf Z}_2$ whose generator preserves the holomorphic volume form.
When we apply the same analysis as above to this particular
gerbe, we find that the resulting theory has all the same one-loop
twisted sectors as $[X/{\bf Z}_2]$.

However, there is a new complication arising in this non-banded
gerbe.  Although all the same one-loop twisted sectors as in a $[X/{\bf Z}_2]$
orbifold arise, none are omitted;
they arise with different multiplicities.
If we let $\xi$ denote the generator of ${\bf Z}_2$,
then the
\begin{displaymath}
{\scriptstyle 1} \square_{1}
\end{displaymath}
one-loop twisted sector of $[X/{\bf Z}_2]$ arises from
\begin{displaymath}
{\scriptstyle \pm 1, \pm i} \square_{ \pm 1, \pm i}
\end{displaymath}
twisted sectors in $[X/{\bf H}]$
{\it i.e.} has multiplicity sixteen,
whereas the
\begin{displaymath}
{\scriptstyle 1} \square_{ \xi}
\end{displaymath}
one-loop twisted sector of $[X/{\bf Z}_2]$ arises from
the $[X/{\bf H}]$ twisted sectors
\begin{displaymath}
{\scriptstyle \pm 1} \square_{ \pm j, \pm k},
\end{displaymath}
and so only has multiplicity eight.
Because $\langle i \rangle$ is not central in ${\bf H}$,
there are no
\begin{displaymath}
{\scriptstyle \pm i} \square_{ \pm j, \pm k}
\end{displaymath}
twisted sectors
in the $[X/{\bf H}]$ orbifold, and so the multiplicity is reduced.
Similarly, the
\begin{displaymath}
{\scriptstyle \xi } \square_{ \xi}
\end{displaymath}
twisted sector of the $[X/{\bf Z}_2]$
orbifold arises from the $[X/{\bf H}]$ twisted sectors
\begin{displaymath}
{\scriptstyle \pm j} \square_{\pm j} \, , \: \: \:
{\scriptstyle \pm k} \square_{\pm k} \, ,
\end{displaymath}
and so again has
multiplicity eight.
                                                                                
Since the $[X/{\bf H}]$ orbifold is explicitly modular-invariant when described
as an ${\bf H}$ orbifold, it must also be modular-invariant
when described in terms of twisted sectors of the $[X/{\bf Z}_2]$
orbifold, and indeed it is straightforward to check that this is
the case.  The $SL(2,{\bf Z})$ action on $[X/{\bf Z}_2]$ one-loop
twisted sectors has two orbits, given by
\begin{displaymath}
\left\{ {\scriptstyle 1} \square_1 \: \right\}
\end{displaymath}
and
\begin{displaymath}
\left\{ {\scriptstyle 1} \square_{\xi} \, , \: {\scriptstyle \xi}
\square_1 \, , \: {\scriptstyle \xi} \square_{\xi} \: \right\} \\
\end{displaymath}
so as multiplicities are constant within each individual
$SL(2,{\bf Z})$ orbit, again we see the theory is modular-invariant.

\subsubsection{Another non-banded gerbe}
                                                                                
Another example of a non-banded gerbe can be obtained as follows.
Consider the nonabelian group $A_4$ \cite[chapter I.5]{lang}
of alternating permutations of
four elements.   One uses the notation
$(a b c \cdots d)$ to indicate a permutation mapping $a$ to $b$,
$b$ to $c$, and so forth, eventually wrapping around to map
$d$ to $a$.  This group has a ${\bf Z}_2 \times {\bf Z}_2$
normal subgroup described by the nontrivial elements
\begin{eqnarray*}
\alpha & \equiv & (1 4) (2 3) \\
\beta & \equiv & (1 3) (2 4) \\
\gamma & \equiv & (1 2) (3 4)
\end{eqnarray*}
The group $A_4$ has a total of twelve elements,
and the three elements of the quotient 
\begin{displaymath}
A_4 / {\bf Z}_2 \times {\bf Z}_2 \cong {\bf Z}_{3}
\end{displaymath}
have representatives
\begin{displaymath}
\begin{array}{c}
\left\{ 1, \alpha, \beta, \gamma \right\} \\
\left\{ (1 2 3), (1 4 2), (2 4 3), (1 3 4) \right\} \\
\left\{ (1 3 2), (1 4 3), (1 2 4), (2 3 4) \right\}
\end{array}
\end{displaymath}
One can form a non-banded ${\bf Z}_2 \times {\bf Z}_2$ gerbe
over an orbifold $[X/{\bf Z}_3]$ as,
$[X/A_4]$, where $A_4$ acts on $X$ by first projecting to ${\bf Z}_3$
and then using the ${\bf Z}_3$ action.
                                                                                
The analysis of strings on this non-banded gerbe proceeds much as before.
We find that every possible one-loop twisted sector of $[X/{\bf Z}_3]$
reappears in $[X/A_4]$ -- no twisted sectors are omitted.
However, the one-loop twisted sectors appear with different multiplicities.
Let us work through a few examples.
Let $\xi$ denote the generator of ${\bf Z}_3$.
The
\begin{displaymath}
{\scriptstyle 1} \square_1
\end{displaymath}
one-loop twisted sectors of the $[X/{\bf Z}_3]$ orbifold
arise from
\begin{displaymath}
{\scriptstyle 1, \alpha, \beta, \gamma} \square_{1, \alpha, \beta, \gamma}
\end{displaymath}
one-loop twisted sectors of the $[X/A_4]$ orbifold, and so have
multiplicity $4^2=16$.  The
\begin{displaymath}
{\scriptstyle 1} \square_{\xi}
\end{displaymath}
one-loop twisted sector of
the $[X/{\bf Z}_3]$ orbifold arises from the
\begin{displaymath}
{\scriptstyle 1} \square_{ (123) } \, , \: \: \:
{\scriptstyle 1} \square_{ (142) } \, , \: \: \:
{\scriptstyle 1} \square_{ (243) } \, , \: \: \:
{\scriptstyle 1} \square_{ (134) }
\end{displaymath}
one-loop twisted sectors of the $[X/A_4]$ orbifold, and so have
multiplicity $4$.  The
\begin{displaymath}
{\scriptstyle \xi} \square_{\xi^2}
\end{displaymath}
one-loop twisted sectors
of the $[X/{\bf Z}_3]$ orbifold arise from the
\begin{displaymath}
{\scriptstyle (123)}\square_{(132)} \, , \: \: \:
{\scriptstyle (142)} \square_{(124)} \, , \: \: \:
{\scriptstyle (243)} \square_{(234)} \, , \: \: \:
{\scriptstyle (134)} \square_{(143)}
\end{displaymath}
one-loop twisted sectors of the $[X/A_4]$ orbifold, and so have
multiplicity $4$.
Proceeding in this fashion, one can show that the
\begin{displaymath}
{\scriptstyle 1} \square_1
\end{displaymath}
twisted
sector of the $[X/{\bf Z}_3]$ orbifold appears with multiplicity
sixteen, but all other one-loop twisted sectors appear with
multiplicity four.
                                                                                
Modular invariance of the theory is guaranteed by its presentation
as an $[X/A_4]$ orbifold, but is also straightforward to check
in terms of $[X/{\bf Z}_3]$ twisted sectors.
The $SL(2,{\bf Z})$ orbits of one-loop twisted sectors of
$[X/{\bf Z}_3]$ are given by
\begin{displaymath}
\left\{ {\scriptstyle 1} \square_1 \: \right\}
\end{displaymath}
\begin{displaymath}
\left\{ {\scriptstyle 1} \square_{\xi} \, , \:
{\scriptstyle \xi} \square_1 \, , \:
{\scriptstyle \xi} \square_{\xi} \, , \:
{\scriptstyle \xi} \square_{\xi^2} \, , \:
{\scriptstyle \xi^2} \square_{\xi} \, , \:
{\scriptstyle \xi^2} \square_{\xi^2} \: \right\}
\end{displaymath}
so again we see that multiplicities are constant on elements of
any given $SL(2,{\bf Z})$ orbit, and so the theory is modular-invariant.

\section{Gauging nonfinite noneffective groups}
\label{nonmincharge}
                                       
In the previous section we discussed several examples of gauged
finite noneffectively-acting groups.  Now, it is also certainly possible
to gauge a nonfinite group acting noneffectively but
with finite stabilizers. 
For example, consider a sigma model on the total space of a principal
$U(1)$ bundle, in which a $U(1)$ is gauged which rotates the fibers
$k$ times instead of once.  By comparison to a gauging which rotates
the fibers only once, rotating the fibers $k$ times means giving
the fields in the worldsheet theory nonminimal $U(1)$ charges.

Since we have seen in examples that gauging a noneffectively-acting
finite group is not equivalent to gauging an effectively-acting group,
we would expect the same to be true of nonfinite groups.
After all, one can describe both as local orbifolds by trivially-acting
groups, so one would expect qualitatively similar behavior.
Thus, in order to be consistent with the observations
of the last section, one expects that a two-dimensional
gauge theory with nonminimal charges must be different from a two-dimensional
gauge theory with minimal charges.

Indeed, that is the case.  Although such two-dimensional gauge
theories are the same perturbatively, they are very different
nonperturbatively.
                                                                                
This fact will play a crucial role in \cite{glsm}, where we will 
study gauged linear sigma models for toric stacks, which 
look like ordinary gauged linear sigma models, but with nonminimal charges.
There, we will explicitly calculate some of the many ways in which
the theories differ -- from different correlation functions to
different R-symmetry anomalies.
                                                                                
Since this physical effect is obscure, let us take a moment to
describe
more carefully the
general reasons why these theories are distinct.  
(We
would like to thank J.~Distler and R.~Plesser for providing the
detailed argument that we review in this section.)
For a different
discussion of two-dimensional gauge theories with fermions of
nonminimal charges, see \cite[section 4]{edold}.  (The discussion
there is most applicable to the present situation when $m \ll M$, in
the notation of that reference.)

To be specific, consider a gauged linear sigma model with a single
$U(1)$ gauge field, and with chiral superfields, all of charge $k$,
with $k>1$.  (Mathematically, this corresponds to a ${\bf Z}_k$ gerbe
on a projective space, as we shall review in \cite{tonyme,glsm}.)
One might argue that this theory should be the same as a theory
with chiral superfields of charge $1$, as follows.
Since instanton number is essentially monopole number,
from Dirac quantization since the electrons have charges a multiple
of $k$, the instantons must have charge a multiple of $1/k$,
and so zero modes of the Higgs fields in a minimal nonzero instanton
background would be sections of ${\cal O}(k/k) = {\cal O}(1)$,
just as in a minimal charge GLSM.  Making the charges nonminimal
has not changed the physics.
In order to recover the physics we have described,
we require the Higgs fields to have charge $k$ while the
instanton numbers are integral, not fractional.
                                                                                
Closer analysis reveals subtleties.
Let us break up the analysis into two separate cases:  first,
the case that the worldsheet is noncompact, second,
that the worldsheet is compact.  For both cases, it will be important
that the worldsheet theory is two-dimensional.
                                                                                
First, the noncompact case.
Since the $\theta$ angle couples to $\mbox{Tr }F$,
we can determine the instanton numbers through the periodicity of
$\theta$.  Suppose we have the physical theory described above,
namely a GLSM with Higgs fields of charge $k$,
plus two more massive fields, of charges $+1$ and $-1$.
In a two-dimensional theory, the $\theta$ angle acts as an electric
field, which can be screened by pair production, and that screening
determines the periodicity of $\theta$.
If the only objects we could pair produce were the Higgs fields
of charge $k$, then the theta angle would have periodicity
$2 \pi k$, and so the instanton numbers would be multiples
of $1/k$.  However, since the space is noncompact, and the
electric field fills the entire space, we can also pair produce
arbitrary numbers of the massive fields, which have charges
$\pm 1$, and so the $\theta$ angle has periodicity $2 \pi$,
so the instantons have integral charges.

We can phrase this more simply as follows.
In a theory with only Higgs fields of charge $k$,
the instanton numbers are multiples of $1/k$, and so the resulting
physics is equivalent to that of a GLSM with minimal charges.
However, if we add other fields of charge $\pm 1$,
then the instanton numbers are integral,
and if those fields become massive, and we work at an energy scale
below that of the masses of the fields, then we have a theory
with Higgs fields of charge $k$, and integral instanton numbers,
giving us the physics that corresponds to a gerbe target.
                                                                                
Thus, we see in the noncompact case that there are two
possible physical theories described by Higgs fields of charge $k$:
one is equivalent to the GLSM with minimal charges,
and the other describes the gerbe.

The analysis for the compact worldsheet case is much shorter.
Strictly speaking, to define the theory nonperturbatively on a
compact space, we must specify, by hand, the bundles that the
Higgs fields couple to.  If the gauge field is described by
a line bundle $L$, then coupling all of the Higgs fields to
$L^{\otimes k}$ is a different prescription from coupling all
of the Higgs fields to $L$.  As a result, the spectrum of zero modes
differs between the two theories, hence correlation functions and
anomalies differ between the two theories,
and so the two physical theories are very different,
as we shall see in examples later.
                                                                                
We shall assume throughout this paper that the worldsheet is
compact, though as we have argued the same subtlety shows up
for noncompact worldsheets.

Again, we shall discuss this matter in much greater detail
in \cite{tonyme,glsm}, but to help whet the reader's appetite, let us
review how this works in a simple example.  Consider the ${\bf C}
{\bf P}^{N-1}$ model, realized as $N$ chiral superfields each of
charge $1$ with respect to a gauged $U(1)$. 
Let us construct a model which we shall denote the $G^k_{-1} {\bf P}^{N-1}$
model (notation to be explained in \cite{glsm}),
or $G {\bf P}^{N-1}$ for brevity,
consisting of $N$ chiral superfields each of charge $k$ with respect
to a single gauged $U(1)$.  Although perturbatively these two two-dimensional
gauge theories are equivalent, nonperturbatively they are distinct.
For example, in the ordinary ${\bf C} {\bf P}^{N-1}$ model,
anomalies break the $U(1)_A$ to a ${\bf Z}_{2N}$ subgroup,
whereas in the $G {\bf P}^{N-1}$ model, anomalies break the
$U(1)_A$ to a ${\bf Z}_{2kN}$ subgroup.  The quantum cohomology
ring of the ordinary ${\bf C} {\bf P}^{N-1}$ model is given by
\begin{displaymath}
{\bf C}[x]/(x^N \: - \: q)
\end{displaymath}
whereas the quantum cohomology ring of the $G {\bf P}^{N-1}$ model
is given by
\begin{displaymath}
{\bf C}[x]/(x^{kN} \: - \: q)
\end{displaymath}
reflecting the fact that A model correlation functions in the two
theories are different.
We shall explore this in much more detail in \cite{glsm}.

\section{Closed string spectra}   \label{spectra}

\subsection{Quotients by finite noneffectively-acting groups}
\label{spectra:noneff}

To compute the massless spectrum of a sigma model on $X$ with a gauged
noneffectively-acting finite group $G$,
one way to proceed is to do the computation formally the same way
as for an effectively-acting finite group:  for each principal $G$-bundle
on $S^1$, we have a branch of the semiclassical moduli space,
and so quantizing that branch we get a sector of the Hilbert space.
In this fashion we are led to
a massless spectrum given by
\begin{displaymath}
\oplus_{[g]} H^*(X^g; {\bf C})^{Z(g)}
\end{displaymath}
where the sum is over conjugacy classes in $G$, and $Z(g)$ is the centralizer
of a given element $g$ representing some conjugacy class.
The inertia stack of $[X/G]$ for $G$ finite and noneffectively-acting
has the same form as for $G$ finite and effectively-acting, namely
\begin{displaymath}
I_{[X/G]} \: = \: \prod_{[g]} [ X^g/Z(g) ]
\end{displaymath}
and so proceeding as before, the massless spectrum is the same as the
de Rham cohomology of the inertia stack.
                                                                                
A skeptic might well argue that this calculation is somewhat naive.
Let us work through a simple example, and examine the details of
the calculation.
                                                                                
Consider for example $[X/{\bf Z}_k]$, where the ${\bf Z}_k$ acts
completely trivially on $X$.  According to the proposed
massless spectrum calculation
above, since ${\bf Z}_k$ is abelian, the Hilbert space should contain
$k$ sectors, and since the ${\bf Z}_k$ acts trivially, $X^g = X$ for all
$g$, so each twisted sector contains a copy of $H^*(X; {\bf C})$.
In other words, according to the calculation above, the massless spectrum
of this orbifold should be $k$ copies of the massless spectrum of a
sigma model on $X$.
                                                                                
The one-loop partition function of this gauged sigma model is given by
\begin{eqnarray*}
Z_{[X/G]} & = & \frac{1}{| {\bf Z}_k | } \sum_{g,h} Z_{g,h} \\
& = & \frac{1}{ | {\bf Z}_k | } | {\bf Z}_k |^2 Z_{1,1} \\
& = & k Z_X
\end{eqnarray*}
just a factor of $k$ times the one-loop partition function for $X$.
                                                                                
Now, ordinarily in quantum field theory, multiplying a partition function
by a constant has no effect on the physics, so a skeptic might argue that
in this case, gauging the trivially-acting ${\bf Z}_k$ should have no effect,
and the massless spectrum should be given by one copy of $H^*(X;{\bf C})$,
not the $k$ copies we obtained above.
However, because this sigma model is ultimately coupled to worldsheet gravity,
we must be more careful.  In a theory coupled to gravity, factors in front
of partition functions cannot be ignored, for the same reasons that
one cannot ignore contributions to a cosmological constant
(see \cite[section 7.3]{pol1} for more details on this).
Thus, the multiplicative factor of $k$ in the one-loop partition function
cannot be consistently ignored.
                                                                                
We can see the effect of such multiplicative factors by closer examination
of the one-loop partition function.  For example, if $X = {\bf R}^d$, then
the one-loop partition function of a sigma model on $X$ can be written
in the form \cite[equ'n~(7.3.8b)]{pol1}:
\begin{displaymath}
Z \: = \: i V_d \int_{F_0} \frac{ d \tau d \overline{\tau} }{ 4 \tau_2 }
\left( 4 \pi^2 \alpha' \tau_2 \right)^{-d/2}
\sum_{i \in {\cal H}^{\perp}} q^{h_i-1} \overline{q}^{ \overline{h}_i-1}
\end{displaymath}
(see the reference for notation) where ${\cal H}^{\perp}$ is
(most of) the closed string Hilbert space.  Multiplying this partition
function by a factor of $k$ looks formally equivalent to increasing
the multiplicity of closed string states by a factor of $k$,
and that is precisely the result we obtained originally for the massless
spectrum.
                                                                                
Another check can be performed by interpreting the one-loop partition function
as a string propagator and counting poles.
For a bosonic string on flat space, the full one-loop partition function
in the regime where $\tau_2 \rightarrow \infty$ has the expansion
\cite[equ'n~(7.3.15)]{pol1}:
\begin{displaymath}
2 \pi i V_{26} \int^{\infty} \frac{d \tau_2}{2 \tau_2}
\left( 4 \pi^2 \alpha' \tau_2\right)^{-13}
\left[ \exp(4 \pi \tau_2) \: + \: 24^2 \: + \: \cdots \right]
\end{displaymath}
The exponential term corresponds to the tachyon in the closed bosonic
string spectrum, the $24^2$ term corresponds to the $24^2$ massless states
of the closed bosonic string
\begin{displaymath}
\alpha_0^{\mu} \overline{\alpha}_0^{\nu} | 0 \rangle
\end{displaymath}
and so forth.
If we were to quotient ${\bf R}^{24}$ by a trivially-acting ${\bf Z}_k$,
the effect would be to multiply this partition function by a factor of $k$.
Then, in this pole expansion, instead of a $24^2$ term, we would have
a $24^2 k$ term, which would indicate $24^2 k$ massless states,
consistent with our calculation of the massless spectrum in the
noneffective orbifold.

A skeptic might nevertheless still want to try to argue that the spectrum of a
trivially-acting
orbifold should only be one copy of the massless spectrum of the cover.
In special cases, namely when the full orbifold group is a central
extension by a trivially-acting group, it is possible to find an
alternative spectrum computation.  If the full orbifold group is nonabelian,
then the one-loop partition function will be proportional to
the partition function of an effectively-acting orbifold (with group
given by the quotient of the full group by the trivially-acting part),
with an $SL(2,{\bf Z})$-orbit of one-loop twisted sectors omitted,
which could be interpreted as modifying the projection operator.
To satisfy such skeptics, we pursue this spectrum calculation program
in section~\ref{falselead}.  Although this direction might sound promising,
ultimately it fails, because the resulting physical theory is
non-unitary.  This is essentially because you cannot consistently
multiply even $SL(2,{\bf Z})$-orbits of one-loop twisted sectors by
zero and get a unitary theory -- although modular invariance is preserved,
multiloop factorization is not.  Unitarity is, in fact, the origin
of the cocycle condition in discrete torsion.  Thus, since this alternative
spectrum calculation leads to nonunitary results, we do not believe
this alternative spectrum calculation is correct.
To help convince skeptics, we work out the details of this false
lead extensively
in section~\ref{falselead}.
                                                                                
Another potential interpretation of the massless spectrum
requires a homomorphism from the trivially-acting group to $U(1)$.
After all, given a banded $G$-gerbe, classified by an element of
$H^2(X,G)$ for $G$ finite, and a homomorphism from $G$ to $U(1)$,
we can construct an element of $H^2(X,U(1))$, which defines a flat
$B$ field.  If such a map arose naturally in these constructions,
then perhaps the correct massless spectrum calculation would be
in terms of an effectively-acting orbifold with a flat $B$ field
background.  However, no such homomorphism arises physically, so far
as we have been able to determine, so this potential massless spectrum
calculation is not well-defined, much less tenable.

A more subtle difficulty, that we have not discussed so far,
involves deformation theory.  The mathematical notion of deformation
theory of a stack encodes only the untwisted sector moduli;
there is no mathematics corresponding to twist field moduli.
Thus, for example, in the orbifold $[X/{\bf Z}_k]$ where
the ${\bf Z}_k$ acts trivially on $X$, the mathematical deformations
are those of $X$.  This would appear to be a problem for the massless
spectrum calculation presented here, as ordinarily the physical
moduli have a geometric understanding as the moduli of the target.
                                                                                
We will briefly discuss this issue later in section~\ref{defthy},
and will discuss the issue much more extensively in \cite{tonyme,glsm}.
Twist fields for trivially-acting group elements can be understood
algebraically, and giving such twist field moduli a vev takes us
into new presentations of abstract CFT's of a form not previously discussed.
We simply have more physical deformations than can be understood
mathematically, and we will be able to see their effects explicitly.
                                                                                
More to the point, we will see explicitly in \cite{glsm} that these
twist field moduli play a critical role in understanding mirror symmetry.
Usual mirror constructions, when applied to quotients by noneffectively-acting
groups, naturally produce the abstract CFT's alluded to above.
Furthermore,
the structure of these abstract CFT's plays a crucial role in understanding
how to generalize Batyrev's mirror construction to stacks.
                                                                                
Since we see these nonmathematical twist field moduli explicitly
giving rise to abstract CFT's, and since we see the same moduli playing
a crucial role in understanding mirror symmetry, we are led to believe
that the proposed calculation of massless spectra in noneffective orbifolds
is correct, and that we have not overcounted states.

\subsection{An instructive false lead on
finite noneffectively-acting
groups}   \label{falselead}
                                                                                
We have just argued that the correct massless spectrum
of an orbifold by a finite non\-effec\-ti\-ve\-ly-acting group should
be computed in formally the same way as for a finite
effectively-acting group:  the Hilbert space has as many sectors
as conjugacy classes of the group, and in each sector,
one takes the part of the cohomology of the fixed-point locus
that is invariant under centralizers.  We have seen how this is
consistent with spectrum calculations based on one-loop partition
function calculations, discussed some alternatives, and also outlined
how this is consistent with deformation theory and mirror symmetry,
topics we shall discuss more extensively later.

We argued in the previous section that multiplicative factors in
orbifold partition functions play a crucial role in checking
state degeneracies, and give a solid test of our massless spectrum
calculation.  To help convince remaining skeptics,
in this subsection we shall see what happens when one ignores those
multplicative factors, and assume that one gets only one copy of untwisted
sector states in noneffective orbifolds, not multiple copies.
This leads to an alternative spectrum calculation, in which omission of
one-loop twisted sectors implies a modified projection operation on the spectrum
of an effectively-acting orbifold. 

We shall see in this subsection that this alternative spectrum calculation
is not consistent, because the resulting physical theories are not
unitary, and moreover this approach does not work in all cases.
To help convince readers that this approach is not fruitful,
and to add support for our proposal, let us work through the details
of this alternative approach, to see in greater detail why it is wrong.

\subsubsection{Basic calculations}
                                                                                
Let us take the attitude that in an orbifold by a noneffectively-acting
finite group, {\it i.e.} a string compactification on a gerbe,
the result should be closely related to the massless spectrum
of an orbifold by an effectively-acting group, given by quotienting out
the noneffectively-acting normal subgroup.
In particular, there should be only one dimension-zero operator,
and omission of some of the one-loop twisted sectors should be interpreted
as modifying the projection operator.  We shall refer to the specific
example of an $[X/D_4]$ orbifold~\ref{ex:d4elliptic},
where the $D_4$ acts by first
projecting to a ${\bf Z}_2 \times {\bf Z}_2$, which acts effectively:
\begin{displaymath}
1 \: \longrightarrow \: {\bf Z}_2 \: \longrightarrow \: D_4 \:
\longrightarrow \: {\bf Z}_2 \times {\bf Z}_2 \:
\longrightarrow \: 1
\end{displaymath}
                                                                                
To understand why the projection operation is modified,
recall that one of the functions of
the one-loop twisted sector sum is to enforce a projection onto
$G$-invariant states in a $G$-orbifold.  Mechanically, summing over
twisted sectors is equivalent to inserting a projection operator
\begin{displaymath}
\frac{1}{|G|} \sum g
\end{displaymath}
in the string propagator that only allows $G$-invariant states to
propagate.  By omitting some of the one-loop twisted sectors,
we no longer have the complete projection operator, so only
a partial projection is enforced.
                                                                                
To see what the projection operator becomes on each $S^1$ twisted
sector, we need to look at the surviving $T^2$ twisted sectors.
Since all twisted sectors of the form $(1 | g)$ for any element
$g \in {\bf Z}_2 \times {\bf Z}_2$ survive, the projection operator
on the untwisted states is the usual one.  Thus, for untwisted states,
we take ${\bf Z}_2 \times {\bf Z}_2$ invariants.
The other $S^1$ twisted sectors are more interesting.
For each $g \in {\bf Z}_2 \times {\bf Z}_2$,
the only surviving $T^2$ twisted sectors involving $g$
are $(1 | g)$ and $(g | g)$.  Thus, in a $g$ twisted sector,
for $g \neq 1$, the projection operator reduces to a projection
onto states invariant under the cyclic subgroup of ${\bf Z}_2 \times
{\bf Z}_2$ generated by $g$.

More generally, given any banded $K$-gerbe $[X/G]$ over an orbifold $[X/H]$
where
\begin{displaymath}
1 \: \longrightarrow \: K \: \longrightarrow \: G \:
\stackrel{\alpha}{\longrightarrow} \: H \: \longrightarrow \: 1
\end{displaymath}
is a central extension involving finite groups,
it is straightforward to see that
all the $[X/H]$ twisted sectors that appear, appear with the same
multiplicity, so that the massless spectrum is given by
\begin{displaymath}
\bigoplus_{ (h) \subset H } H^*\left( [ X^h / Z'(h) ]; {\bf C} \right)
\end{displaymath}
where the sum is over conjugacy classes in $H$, and
$Z'(h) = \alpha( Z( \alpha^{-1}(h) ))$.
                                                                                
In the present case, since ${\bf Z}_2 \times {\bf Z}_2$ acts
freely, the massless spectrum would be just the ${\bf Z}_2 \times {\bf Z}_2$
invariant part of the cohomology of the elliptic curve.
In the language of the paragraph above, whenever $H$ acts freely on $X$,
the massless spectrum would just be the $H$-invariant part of the
massless spectrum of $X$.

\subsubsection{An example}   \label{ex:falselead}

Next, let us consider a specific example.
Consider the ${\bf Z}_2 \times {\bf Z}_2$ action on
$T^6$ in which each ${\bf Z}_2$ flips the signs of
two of the three complex coordinates, as in \cite{vafaed}.
We can define an action of the group $D_4$ on $T^6$,
where $D_4$ is the nontrivial ${\bf Z}_2$ extension of ${\bf Z}_2 \times
{\bf Z}_2$ discussed above, in which $D_4$ acts on $T^6$ by
first projecting to ${\bf Z}_2 \times {\bf Z}_2$ and then
${\bf Z}_2 \times {\bf Z}_2$ acts on $T^6$ as just discussed.
The ${\bf Z}_2$ subgroup of $D_4$ acts trivially, so this corresponds
to a sigma model on a ${\bf Z}_2$ gerbe over the stack
$[ T^6 / {\bf Z}_2 \times {\bf Z}_2 ]$.
This gerbe is Calabi-Yau, and can be shown to be nontrivial.

The physical analysis of closed strings on this gerbe proceeds just
as before.  Recall from \cite{vafaed} that the Hodge diamond of massless
closed string states of the
original $[T^6/ {\bf Z}_2 \times {\bf Z}_2 ]$ is given by
\begin{displaymath}
\begin{array}{ccccccc}
 & & & 1 & & & \\
 & & 0 & & 0 & & \\
 & 0 & & 51 & & 0 & \\
1 & & 3 & & 3 & & 1 \\
 & 0 & & 51 & & 0 & \\
 & & 0 & & 0 & & \\
 & & & 1 & & & \end{array}
\end{displaymath}
where
\begin{displaymath}
\begin{array}{ccccccc}
 & & & 1 & & & \\
 & & 0 & & 0 & & \\
 & 0 & & 3 & & 0 & \\
1 & & 3 & & 3 & & 1 \\
 & 0 & & 3 & & 0 & \\
 & & 0 & & 0 & & \\
 & & & 1 & & & \end{array}
\end{displaymath}
states are ${\bf Z}_2 \times {\bf Z}_2$-invariant untwisted sector
states, and the remaining states are $3 \cdot 16$ copies
(one for each nontrivial element of ${\bf Z}_2 \times {\bf Z}_2$,
and one for each fixed point locus under a fixed element) of
the ${\bf Z}_2 \times {\bf Z}_2$-invariant elements of
\begin{displaymath}
\begin{array}{ccccccc}
 & & & 0 & & & \\
 & & 0 & & 0 & & \\
 & 0 & & 1 & & 0 & \\
0 & & 1 & & 1 & & 0 \\
 & 0 & & 1 & & 0 & \\
 & & 0 & & 0 & & \\
 & & & 0 & & & \end{array}
\end{displaymath}
which is to say, $3 \cdot 16$ copies of
\begin{displaymath}
\begin{array}{ccccccc}
 & & & 0 & & & \\
 & & 0 & & 0 & & \\
 & 0 & & 1 & & 0 & \\
0 & & 0 & & 0 & & 0 \\
 & 0 & & 1 & & 0 & \\
 & & 0 & & 0 & & \\
 & & & 0 & & & \end{array}
\end{displaymath}
The spectrum calculation for the ${\bf Z}_2$ gerbe
$[ T^6 / D_4 ]$ over $[ T^6 / {\bf Z}_2 \times {\bf Z}_2 ]$
is almost identical, except that now in the twisted sectors, we only
project onto states invariant under the subgroup generated by the
group element associated with the twisted sector, not onto states
invariant under all of ${\bf Z}_2 \times {\bf Z}_2$.
Thus, the Hodge diamond of massless states on the gerbe is given by
a sum of
\begin{displaymath}
\begin{array}{ccccccc}
 & & & 1 & & & \\
 & & 0 & & 0 & & \\
 & 0 & & 3 & & 0 & \\
1 & & 3 & & 3 & & 1 \\
 & 0 & & 3 & & 0 & \\
 & & 0 & & 0 & & \\
 & & & 1 & & & \end{array}
\end{displaymath}
states from the untwisted sector, invariant under the entire
${\bf Z}_2 \times {\bf Z}_2$, plus $3 \cdot 16$ copies of
\begin{displaymath}
\begin{array}{ccccccc}
 & & & 0 & & & \\
 & & 0 & & 0 & & \\
 & 0 & & 1 & & 0 & \\
0 & & 1 & & 1 & & 0 \\
 & 0 & & 1 & & 0 & \\
 & & 0 & & 0 & & \\
 & & & 0 & & & \end{array}
\end{displaymath}
which are twisted sector states,
invariant under the relevant subgroup of ${\bf Z}_2 \times {\bf Z}_2$.
Thus, the Hodge diamond of massless states on the gerbe
$[T^6/D_4]$ is given by
\begin{displaymath}
\begin{array}{ccccccc}
 & & & 1 & & & \\
 & & 0 & & 0 & & \\
 & 0 & & 51 & & 0 & \\
1 & & 51 & & 51 & & 1 \\
 & 0 & & 51 & & 0 & \\
 & & 0 & & 0 & & \\
 & & & 1 & & & \end{array}
\end{displaymath}
We see that $H^{1,1}$ and $H^{2,2}$ of the gerbe $[T^6/D_4]$
are identical to $H^{1,1}$ and $H^{2,2}$ of the underlying orbifold
$[T^6/{\bf Z}_2 \times {\bf Z}_2]$, but $H^{1,2}$ and $H^{2,1}$ of
the gerbe are significantly larger -- the gerbe has $48$ extra
complex structure deformations beyond those possessed by the
underlying orbifold, according to this proposed massless spectrum calculation.

We shall see in the next subsection that this proposed alternative
massless spectrum
calculation fails the test of unitarity.

\subsubsection{Unitarity fails in the alternate interpretation}
                                                                                
Now that we have examined one-loop twisted sectors,
let us take a moment to consider higher-loop
twisted sectors.
In particular, we will show
that the one-loop $[T^2/{\bf Z}_2 \times {\bf Z}_2]$
twisted sectors that are
`forbidden' in the $[T^2/D_4]$ orbifold can reappear at higher string
loop order, which is a sign of nonunitarity in this
alternate interpretation of noneffective orbifolds.
We take this result as another indication that our original interpretation
of noneffective orbifolds and their spectra is correct.
                                                                                
For example, consider a two-loop twisted sector in
$[T^2/D_4]$.  It is defined by four group elements $g_1$, $h_1$,
$g_2$, $h_2$ which must obey the relation
\begin{displaymath}
h_1 g_1^{-1} h_1^{-1} g_1 \: = \: g_2^{-1} h_2 g_2 h_2^{-1}
\end{displaymath}
in the conventions of \cite[section 4.3.2]{dt3},
just as the two group elements defining a one-loop twisted sector
must obey the constraint that they commute.
When both sides of the equation above are separately equal to the
identity, the two-loop diagram factors through the identity operator,
and can degenerate into
a pair of one-loop twisted sectors joined by a long thin handle.
In the present case of a $[T^2/D_4]$ orbifold,
consider the case that
\begin{eqnarray*}
g_1 & = & a \\
h_1 & = & ab \\
g_2 & = & a \\
h_2 & = & b
\end{eqnarray*}
These four group elements satisfy the condition above,
and so define a two-loop twisted sector.
Moreover, these four group elements obey the condition
\begin{displaymath}
h_1 g_1^{-1} h_1^{-1} g_1 \: = \: z \: = \:
 g_2^{-1} h_2 g_2 h_2^{-1}
\end{displaymath}
Since $z$ acts trivially on $T^2$, in the target space this two-loop
diagram appears factorizable -- it looks like a product of two
one-loop diagrams.  The one-loop factors, however, are
`forbidden' diagrams -- since $a$ and $b$ do not commute
as elements of $D_4$,
there is no $(a|b)$ one-loop twisted sector,
and similarly there is no $(a|ab)$ one-loop twisted sector,
despite the fact that both reappear inside this two-loop diagram.
                                                                                
This lack of factorization is a signal of failure of unitarity of the
target space theory.  Recall that 
the optical theorem \cite[section 3.6]{weinberg1},
an immediate consequence of unitarity of the
S-matrix, says that the imaginary parts of a scattering amplitude
can be obtained by cutting the diagram in half,
multiplying the amplitudes for the two separate halves,
and integrating over intermediate momenta.
In a little more detail, following \cite[section 3.6]{weinberg1},
if we write S-matrix elements as
\begin{displaymath}
S_{\beta \alpha} \: = \: \delta(\beta - \alpha) \: - \:
2 \pi i \delta(p_{\beta} - p_{\alpha}) M_{\beta \alpha}
\end{displaymath}
(see the reference for notation)
then the relation $\sum_{\beta} S_{\beta \gamma}^* S_{\beta \alpha} =
\delta(\gamma - \alpha)$ implies that
\begin{displaymath}
\mbox{Im } M_{\alpha \alpha} \: = \: - \pi \sum_{\beta}
\delta(p_{\beta} - p_{\alpha}) | M_{\beta \alpha} |^2
\end{displaymath}
In the present case, since the one-loop diagrams vanish,
in a unitary theory we would expect that the two-loop diagram has to vanish,
but that is not what we found -- the two-loop diagram is nonvanishing,
whereas the one-loop diagram vanishes.  Thus, we appear to violate
the optical theorem, and hence violate unitarity
(unless all relevant scattering amplitudes have no imaginary part,
which seems extremely unlikely).
More generally, the optical theorem in the target-space theory
is the reason why factorization of higher-loop amplitudes is necessary
for unitarity.
                                                                                
In passing, note that this same argument does {\it not} apply when we
calculate the massless spectrum using the methods we support.
If we do not omit noneffectively-acting twist fields, if the
massless spectrum contains multiple dimension zero operators
in different (noneffective) twisted sectors, then the two-loop diagram
above does not factorize on the identity, but rather on a dimension-zero
twist field.  In this case, the two one-loop diagrams appearing on either
side of the cut are not one-loop vacuum diagrams, but rather contain a
(noneffective) twist field insertion, and so need not vanish, thereby
preventing a contradiction with unitarity.
                                                                                
There is an another way to see that unitarity is broken in this alternative
interpretation of the noneffective orbifold, based on a difficulty with the
fusion rules.
Consider a gerbe over an orbifold, with extra twisted sector
states.  In particular, consider a gerbe over a ${\bf Z}_2 \times {\bf Z}_2$
orbifold, such as our noneffective $D_4$ orbifold example.
If we have a state in the $a$ twisted sector that is not invariant
under $b$, where $a$ and $b$ generate ${\bf Z}_2 \times {\bf Z}_2$,
then consider the product of that state with another $a$ sector
state that is invariant under $b$.  Since $a^2=1$, the result is an untwisted
sector state that's not invariant under $b$ -- which cannot be allowed!
(Unless it is the zero state.)
The algebra does not close, so the theory does not make sense,
as the operator products are not well-defined.

Although lack of unitarity is not necessarily
completely fatal (for example, noncommutative field theories are
often nonunitary \cite{gomis}), in the present case we find this explicit
failure of unitarity to be suggestive, and in light of other
arguments presented earlier, we do not believe this alternative calculation
of massless spectra to be correct.
Thus, we are led to believe
that the correct massless spectrum of a noneffective orbifold 
has as many sectors in the Hilbert space as conjugacy classes in
the group, even if some of the group elements act trivially.

\subsection{Quantum symmetries in noneffective orbifolds}  \label{quantumsymm:noneff}

We have argued that the Hilbert space of a noneffective orbifold
should be computed in the same form as that of an effective orbifold.
Recall that an effective orbifold has a `quantum symmetry'
\cite{vafaqs}.
In an abelian effective orbifold $[X/G]$, for $G$ finite,
the quantum symmetry is $G$, and gives phases to twisted sectors.
                                                                                
Because of the form of our result,
noneffective orbifolds also trivially possess the same quantum
symmetry.  For example, in an abelian orbifold $[X/G]$ where $G$ is finite
and acts trivially, there is a quantum symmetry $G$ which multiplies
the twisted sectors by phases.
                                                                                
In effective abelian orbifolds, orbifolding the orbifold by the
quantum symmetry restores the original theory.  The same arguments
used to establish 
this fact (see {\it e.g.} \cite[section 8.5]{ginsparg}) can now
be trivially extended to noneffective orbifolds, where one can
easily see the same result is obtained.
                                                                                
Now, suppose ${\cal C}$ is a CFT with a ${\bf Z}_n$ action,
so that the orbifold CFT ${\cal C}' \equiv [ {\cal C} / {\bf Z}_n]$
has a ${\bf Z}_n$ quantum symmetry.  Suppose we now orbifold
${\cal C}'$ by ${\bf Z}_{kn}$ where ${\bf Z}_{kn}$ acts
(noneffectively) on ${\cal C}'$ by first projecting to ${\bf Z}_n$,
\begin{displaymath}
1 \: \longrightarrow \: {\bf Z}_k \: \longrightarrow \:
{\bf Z}_{kn} \: \longrightarrow \: {\bf Z}_n \: \longrightarrow \: 1
\end{displaymath}
and then letting the ${\bf Z}_n$ act on the quantum symmetry.
A natural guess is that the orbifold $[{\cal C}'/{\bf Z}_{kn}]$
should give the same physical theory as the orbifold
of the original CFT ${\cal C}$ by a trivially-acting ${\bf Z}_k$.
Let us take a moment to see that explicitly, 
following \cite[section 8.5]{ginsparg}.
                                                                                
First, let us recall why the ${\bf Z}_n$ orbifold of a $[ {\cal C}/{\bf Z}_n]$
orbifold is again the original CFT ${\cal C}$.
Let the generator of (either) ${\bf Z}_n$ be denoted $g$,
let the one-loop twisted sector with boundaries $g^a$, $g^b$ in
the $[ {\cal C}/{\bf Z}_n]$ orbifold be denoted 
\begin{displaymath}
{\scriptstyle a} \square_b,
\end{displaymath}
and the one-loop twisted sector with analogous boundary conditions
in the ${\bf Z}_n$ orbifold of the $[ {\cal C}/{\bf Z}_n]$ orbifold be
denoted \begin{displaymath}
{\scriptstyle a} \square_b\,{'}.
\end{displaymath}
Let $\xi$ be the generator of the $n$th roots of
unity.  Then, it is straightforward to show that
\begin{displaymath}
{\scriptstyle a} \square_{b}\,{'} \: = \: \frac{1}{n} \sum_{c,d} \xi^{ac} \xi^{bd}
\left( {\scriptstyle c} \square_d \right)
\end{displaymath}
so that the complete one-loop partition function is
\begin{eqnarray*}
Z'' & = & \frac{1}{n} \sum_{a,b} \left( {\scriptstyle a} \square_b\,{'} \right) \\
& = &  \frac{1}{n^2} \sum_{a,b} \sum_{c,d} \xi^{ac} \xi^{bd}
\left(  {\scriptstyle c} \square_d \right) \\
& = & \frac{1}{n^2} \sum_{c,d} n^2 \delta_{c,0} \delta_{d,0}
\left(  {\scriptstyle c} \square_d \right) \\
& = & {\scriptstyle 0} \square_0
\end{eqnarray*}
which is the
one-loop partition function for the original CFT ${\cal C}$.
                                                                                
Now, let us repeat this analysis for a ${\bf Z}_{kn}$ orbifold of
$[ {\cal C}/ {\bf Z}_n]$, where the ${\bf Z}_k$ kernel acts trivially
and the ${\bf Z}_n$ projection acts as the quantum symmetry.
Using indices $i,j \in \{ 0, 1, \cdots, kn-1\}$,
and the fact that each twisted sector in this orbifold will be the
same as a twisted sector in the ${\bf Z}_n$ orbifold of
$[ {\cal C}/{\bf Z}_n]$, we have that
\begin{displaymath}
{\scriptstyle i} \square_j\,{'}
 \: = \: \frac{1}{n} \sum_{a,b=0}^{n-1} \xi^{a [i/k]}
\xi^{b [j/k]} \left( {\scriptstyle a} \square_b \right)
\end{displaymath}
Thus, the full one-loop partition function of the final orbifold is
given by
\begin{eqnarray*}
Z'' & = & \frac{1}{kn} \sum_{i,j=0}^{kn-1} \left( {\scriptstyle i} \square_j\,{'}
\right) \\
& = & \frac{1}{kn^2} \sum_{i,j=0}^{kn-1} \sum_{a,b=0}^{n-1}
\xi^{a [i/k]}
\xi^{b [j/k]}  \left( {\scriptstyle a} \square_b \right) \\
& = & \frac{1}{kn^2} \sum_{a,b=0}^{n-1} (kn)^2 \delta_{a,0} \delta_{b,0}
 \left( {\scriptstyle a} \square_b \right) \\
& = & k \left( {\scriptstyle 0} \square_0 \right)
\end{eqnarray*}
the same as the one-loop partition function of
the orbifold $[ {\cal C}/{\bf Z}_k]$ where the ${\bf Z}_k$ acts trivially.
Thus, we have confirmation of our conjecture.
                                                                                
We have only described one-loop partition functions, but the calculation
can be repeated at arbitrary genus.
It is straightforward to compute
that the $g$-loop partition function of the ${\bf Z}_{kn}$ orbifold
of $[ {\cal C}/{\bf Z}_n]$ is given by
\begin{displaymath}
\frac{1}{(kn)^g} \frac{1}{n^g} (kn)^{2g} \: = \: k^g
\end{displaymath}
times the $g$-loop partition function
of ${\cal C}$, which is the same as the $g$-loop partition function
of the orbifold $[ {\cal C}/{\bf Z}_k]$ for a trivially-acting
${\bf Z}_k$.

\section{CFT and trivial group actions}    \label{trivgerbesection}

In this section we collect some remarks on
gauging a $G$-action on $X$ where all of $G$ acts trivially.
We will assume that $G$ is finite.     
                                                                           
In order to make contact with our upcoming work
\cite{tonyme,glsm}, let us note that mathematically, 
a quotient by a $G$-action in which all of $G$ acts trivially
is the same as the trivial $G$-gerbe.

\subsection{Trivial group actions and product CFT's} \label{trivgcft}
                                                                                
We have argued that the massless
spectrum of a global quotient by a
finite noneffectively-acting group should be computed in exactly
the same fashion as an effectively-acting group, with one sector
of the Hilbert space for each conjugacy class in $G$, and so forth.
In the case of a trivial gerbe presented as above,
this means that there are as many twisted sectors as conjugacy classes
of $G$, and each twisted sector is additively a copy of the untwisted
sector, with a dimension zero operator in each twisted sector
corresponding to the identity of the untwisted sector.
Furthermore, on the basis of quantum numbers it is clear that
a state in any given twisted sector can be obtained from its counterpart
in the untwisted sector by acting on the untwisted state with the
dimension zero twist field.
                                                                                
Put more simply, the massless spectrum, both additively and in its
product structures, looks like the tensor product of the CFT for
the underlying Calabi-Yau $X$ and the CFT for the orbifold $[\mbox{point}/G]$.
The spectrum of the latter orbifold contains only dimension zero operators,
one for each conjugacy class of $G$, and multiplying them by the
identity operator in the CFT for $X$ generates the dimension zero twist
fields in the CFT of $[X/G]$.
                                                                                
This similarity with the tensor product extends to one-loop
partition functions.
Recall the one-loop partition function for the trivial gerbe on $X$
is given by
\begin{displaymath}
|G| Z(X)
\end{displaymath}
By comparison, in conventions in which the partition function for
a sigma model on a point is $1$, the one-loop partition function for the
orbifold $[\mbox{point}/G]$ is given by $|G|$.  The one-loop partition
function of the tensor product is the product of the partition functions
for the separate theories, so we see that the partition function
for the tensor product of a sigma model on $X$ ($Z(X)$)
and the orbifold $[\mbox{point}/G]$ is given by $|G|Z(X)$, matching
the one-loop partition function for the trivial gerbe.

On the basis of the massless spectrum, correlation functions,
and the one-loop partition functions,
we claim that physically the CFT corresponding to an orbifold
$[X/G]$ by a global trivial $G$ action
on a Calabi-Yau $X$ is the same as the tensor product of the CFT
${\cal C}_X$ corresponding to $X$ and
the CFT ${\cal C}_G$ corresponding to the orbifold $[\mbox{point}/G]$,
{\it i.e.}
\begin{displaymath}
{\cal C}_{ [X/G] } \: \cong \: {\cal C}_X \otimes {\cal C}_G.
\end{displaymath}

This claim about physics has a mathematical counterpart.
Mathematically, a trivial $G$-gerbe $[X/G]$ over a manifold $X$ can be
expressed as the product $X \times BG$ of stacks,
{\it i.e.}
\begin{displaymath}
[X/G] \: \cong \: X \times BG
\end{displaymath}
where $BG = [\mbox{point}/G]$.
It is straightforward to check this statement at the level of incoming
maps.  A map from a manifold $Y$ into the trivial $G$-gerbe $[X/G]$ is a pair
consisting of a principal $G$-bundle $E$ over $Y$, together
with a $G$-equivariant map $f: E \rightarrow X$.
Since $G$ acts trivially on $X$, $f$ is equivalent to a map
$f': Y \rightarrow X$.  Thus, our map from $Y$ into $[X/G]$ is the
same as a principal $G$-bundle $E$ over $Y$ together with a map
$f': Y \rightarrow X$.  However, that pair also specifies a map
from $Y$ into $X \times BG$.  The relevance of the map $f': Y \rightarrow X$
is clear, and since $BG = [\mbox{point}/G]$,
a map $Y \rightarrow BG$ is just\footnote{This entertaining fact, an immediate
consequence of the definition, makes the stack $[\mbox{point}/G]$ behave
analogously to the classifying space for $G$, and is a reason for the similar
notation and the similar name (classifying stack). } a principal $G$-bundle $E$ over $Y$.
                             
Intuitively, if we try to compare gerbes to fiber bundles,
then $BG$ is the analogue of the fiber of a $G$-gerbe.
A trivial $G$-gerbe over $X$ is the product $X \times BG$.
Also,
all $G$-gerbes over $X$ look locally like $X \times BG$,
though only the trivial gerbe has that form globally.
                                                                                
For $X$ and $Y$ Calabi-Yau spaces, the CFT of $X \times Y$ is the same
as the tensor product of the CFT's corresponding to $X$ and $Y$,
so it is very natural for the CFT of $X \times BG$ to be the tensor product
of the CFT's for $X$ and $BG$.

\subsection{$[\mbox{point}/{\bf Z}_k]$ and finite-group physics}
\label{finitegroupfieldintro}
                                                                                
Consider the CFT defined by the ${\bf Z}_k$-orbifold of a point,
$[\mbox{point}/{\bf Z}_k]$.  The corresponding massless spectrum is
generated by a single twist field $\xi$, as seen in
section~\ref{spectra:noneff},
and because of selection
rules for noneffective orbifolds discussed in
section~\ref{quantumsymm:noneff},
correlation functions $\langle \xi^n \rangle$ vanish unless $n$ is a
multiple of $k$. 
                                                                                
The same result can be obtained from a slightly different-looking setup.
Consider a physical theory defined by a ${\bf Z}_k$-valued field
$\phi$, ${\bf Z}_k$-valued in the sense that it takes values in
the $k$th roots of unity.  One can build a very trivial QFT of this
field:  the path integral measure is just a sum over the $k$ possible
values of $\phi$, the action vanishes identically,
and correlation functions are just simple
statistical measures:
\begin{displaymath}
\langle \phi^n \rangle \: = \: \sum_k \phi^k
\end{displaymath}
Since the path integral measure is just a sum over $k$th roots of unity,
the correlation functions in this trivial theory vanish unless $n$ is
a multiple of $k$.
                                                                                
Since this trivial QFT has the same fields as the CFT
$[\mbox{point}/{\bf Z}_k]$, and those fields have the same correlation
functions, we claim that the trivially-acting orbifold of a point is
isomorphic to this trivial theory of a ${\bf Z}_k$-valued field.
                                                                                
This observation will play an important role later in \cite{tonyme}
when we study deformation theory of stacks -- we will use fields
valued in roots of unity to make some physical deformations without
mathematical counterparts explicit, and so verify their existence.
We shall also, independently, find such fields valued in roots of unity
occurring in \cite{glsm} when we study mirrors to stacks.

\subsection{Trivial group actions versus disconnected targets}
                                                                                
While discussing orbifolds by trivially-acting
${\bf Z}_k$'s, {\it i.e.} trivial ${\bf Z}_k$ gerbes, let us take a moment
to compare their physics to that of sigma models with target space
$k$ disjoint copies of a manifold $X$.
                                                                                
The partition functions of these two theories match.
As already discussed, the $g$-loop partition function of the
orbifold of $X$ by a trivially-acting ${\bf Z}_k$ is
$k^{2g}/k^g = k^g$ times the $g$-loop partition 
function of a sigma model on $X$.
This is also true for the $g$-loop partition function of a sigma
model on $k$ disjoint copies of $X$.  To see this, note that such a sigma
model has $k$ times as many states as a sigma model on $X$, given by
the $k$-fold tensor product of the states of a sigma model on $X$.
Since in a genus $g$ partition function one has states propagating on
each of $g$ loops, the result is that a genus $g$ partition function
when the target is $k$ copies of $X$ should be $k^g$ times the partition
function for $X$, matching the $g$-loop partition function of the trivial
gerbe.
                                         
Similarly, the massless spectra are also the same.                                       
The massless spectrum of the $[X/{\bf Z}_k]$ orbifold is the sum of
$k$ copies of the cohomology of $X$.  Thus, for example, it contains
$k$ dimension zero operators, in each of $k$ twisted sectors.
                                                                                
The massless spectrum of the disjoint union of $k$ copies of $X$ is the
direct sum of $k$ copies of the cohomology of $X$.  In other words,
a state in this sigma model is a $k$-tuple of states in a sigma model on $X$.
Just as in the orbifold $[X/{\bf Z}_k]$, in this theory there are $k$
dimension zero operators, corresponding to the fact that the cohomology
of the disjoint union of $k$ copies of $X$ is dimension $k$ in degree zero.

In fact, we believe these conformal field theories are the same,
and more generally, we believe that noneffective gaugings are
at least often described by the same conformal field theories as
disjoint unions of spaces.  
Furthermore, this identification solves an important technical problem
involving cluster decomposition in these theories:  having multiple
dimension zero operators violates cluster decomposition, but conformal
field theories describing disjoint unions of spaces violate
cluster decomposition in the mildest possible way, causing no other
physical inconsistencies.
We shall discuss these issues in much greater detail in
\cite{clusterdecomp}, and defer further discussion to that work.

\section{Deformation theory issues}   \label{defthy}

One important issue we have not addressed so far concerns the interpretation
of the marginal operators in the CFT's we have described.
In typical examples, there are more marginal operators than there are
geometric moduli, so naturally one must ask, what does it mean to deform
along those directions?
                                                                                
For example, for a trivial ${\bf Z}_k$ orbifold of a space $X$,
{\it i.e.} the trivial gerbe $[X/{\bf Z}_k] = X \times B {\bf Z}_k$,
we have argued in section~\ref{spectra:noneff} that the massless spectrum
is $k$ copies of the cohomology of $X$.  However, the only obvious geometric
deformations are just deformations of $X$, the original untwisted sector
moduli.  What does it mean to deform along the other $k-1$
marginal operators?

This physical puzzle has a mathematical analogue.
The mathematical infinitesimal
moduli of the algebraic stack corresponding to this gerbe
contain only one copy of the moduli of $X$, not $k$ copies.
Again, we have a mismatch.  
Moreover, in a sigma 
model on a smooth manifold $X$, the physical moduli match
mathematical moduli, so the present mismatch is a potential problem,
just as for quotients by effectively-acting finite groups.

One conceivable answer is that the `extra' $k-1$ marginal operators
are obstructed.  However, it is easy to check that since they differ
from the untwisted sector operators merely by the addition of a twist
field associated to a trivially-acting group element, there is no way
to get nonvanishing correlation functions involving these operators
unless there are some nonvanishing correlation functions among the
original untwisted sector operators.  If the untwisted sector operators
are truly marginal, describing unobstructed moduli, then the twisted
sector operators must also be unobstructed.

Another conceivable answer is that there is some subtle 
problem with our massless
spectrum calculation.  After all, the marginal operators are, by definition,
part of the massless spectrum, so if we have miscomputed the massless
spectrum, then we may also have miscomputed the number of marginal
operators.  However, we have performed extensive independent tests
of the massless spectrum calculation, and we do not believe that it is
in error.

This mismatch of deformations is part of a larger issue involving
how physical deformations and mathematical moduli of stacks are related.
We will discuss this matter extensively in \cite{tonyme}.
As we shall discuss in that reference,
in general terms our resolution of such mismatches is that
the mathematical moduli correspond to deformations of the stack
which result in (weakly-coupled) physical theories with well-behaved 
mathematical
interpretations.  The `extra' physical moduli result in physical
theories which do not appear to have clean mathematical interpretations.
We believe this claim because we are able to very explicitly describe
and manipulate the theories that result from such deformations,
as we shall outline below.         
             
In the case of the trivial ${\bf Z}_k$ orbifold of $X$, as described
above, the untwisted sector moduli merely deform the covering space $X$,
an operation which has a clean mathematical understanding.
Giving a vev to twisted sector moduli has a different effect in
conformal perturbation theory:  formally, if we try to insert an
exponential of a second descendant of a twisted sector marginal operator
in correlation functions, defined by its Taylor expansion,
then by the usual selection rules many of the terms in the Taylor expansion
drop out.  This formal operation is no longer anything as clean or simple
as merely an ordinary geometric deformation of $X$.
Rather, one appears to get a new and different family of conformal field theories.

Another example should make the analysis clearer.
Begin with a Landau-Ginzburg model corresponding to a Calabi-Yau hypersurface,
so that the superpotential is the hypersurface polynomial.
As is well-known,
a marginal deformation of the theory corresponds to a deformation
of that superpotential by terms which do not change the degree of
homogeneity of the polynomial.
                                                                                
Now, construct a trivial ${\bf Z}_k$ orbifold of that Landau-Ginzburg
model.  As discussed previously in section~\ref{finitegroupfieldintro},
this is equivalent to adding a field $\Upsilon$ that takes values
in $k$th roots of unity.  According to our massless spectrum calculation,
we now have $k$ times as many moduli in the physical theory as before,
given by multiplying any vertex operator corresponding to a modulus
of the original theory by a power of $\Upsilon$.
Giving a vev to such a twisted sector modulus is equivalent to
adding a term to the Landau-Ginzburg superpotential which has a factor
of $\Upsilon$ to some power, since such terms are just supersymmetry
transformations of the relevant vertex operator.
For example, in \cite{glsm} we shall see Landau-Ginzburg superpotentials
of the form
\begin{displaymath}
W \: = \: x_1^5 \: + \: \cdots \: + \: x_5^5 \: + \:
\Upsilon \psi x_1 x_2 x_3 x_4 x_5
\end{displaymath}
where the $x_i$ are chiral superfields, $\psi$ is a complex number,
and $\Upsilon$ takes values in roots of unity, which are summed over
in the path integral measure.
Thus, we can see these new physical deformations very explicitly,
as {\it e.g.} Landau-Ginzburg superpotential terms with factors
of $\Upsilon$, the field valued in roots of unity.

In principle, we can interpret this in the same way as in the
previous example, adding a formal
exponential of a second descendant of the twisted sector modulus,
and because of the usual selection rules, many of the terms will drop out.
This is an equivalent description.  By working with fields valued in roots
of unity, however, we have a more algebraic description of the
new conformal field theories, something much easier to work with than
the description provided directly by conformal perturbation theory.

The reader might ask why one cannot do the same for twist fields
in effective orbifolds. 
Ordinarily, giving a vev to a twist field is somewhat messy,
as the twist field introduces a branch cut, and as this changes the moding
of worldsheet fields, the resulting marginal operators are no longer
quite so simple to express.  In the present case, however,
since the twist field in question corresponds to a group element that
acts trivially, the moding of worldsheet fields does not change,
and so an algebraic description of the process of giving a vev to a
twist field, as we have outlined above, becomes possible.
                                                                                
We will return to these abstract CFT's and study them more extensively
in \cite{glsm}, where they will be derived from a completely different
direction.  Here, we have derived Landau-Ginzburg models with fields
taking values in roots of unity from considering physical deformations
of noneffective orbifold theories.  In \cite{glsm} we will find that the
same sort of Landau-Ginzburg theories appear when one builds mirrors
to stacks.  The fact that we are seeing these same physical theories
appear in a different context is an excellent check that our
analysis is consistent.

\section{D-branes in noneffective orbifolds}   \label{Dbranenoneff}
                                                                                
Earlier in section~\ref{spectra} we 
argued that the
closed string massless spectrum in a noneffective orbifold
should have exactly the same general form
as that for effectively-acting finite groups.
For example,
even for an orbifold $[X/{\bf Z}_n]$ where the ${\bf Z}_n$ acts
completely trivially, there should still be $n$ distinct twisted
sectors in the massless spectrum, although additively each sector
is identical to the untwisted sector.
                                                                                
\begin{figure}
\centerline{\psfig{file=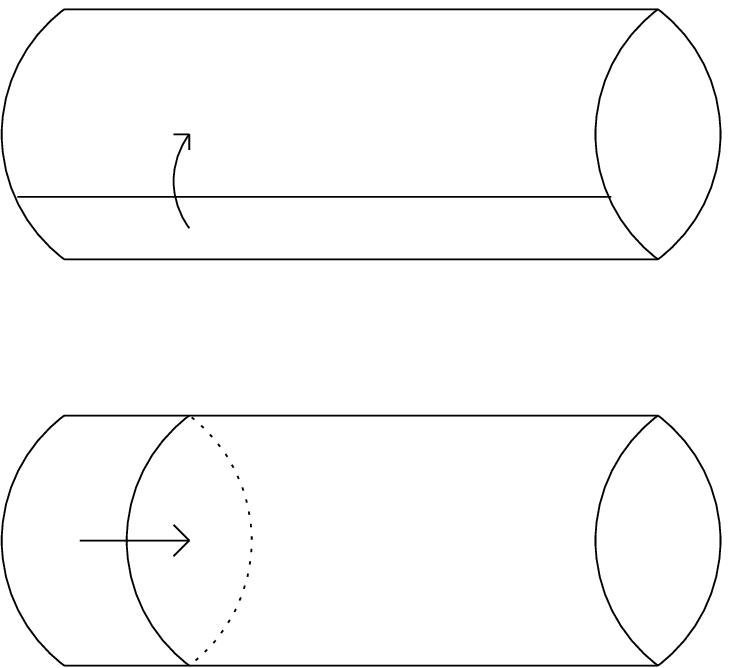,width=2.5in}}
\caption{\label{c1} An annulus diagram, interpreted in two ways}
\end{figure}

For open strings in noneffective orbifolds,
although the group acts trivially on the base space,
it can still act nontrivially on the Chan-Paton factors.
Checking this statement requires verifying the Cardy condition,
as discussed for orbifolds in {\it e.g.} \cite{bcr}.
Recall that Cardy's condition \cite{cardy,cardylew,lew}
amounts to the statement that
the physics of an
annulus diagram (see figure~\ref{c1}) should be independent of whether we
interpret it as an open string propagating
at one-loop or a closed string propagating at tree level
between boundary states.

\begin{figure}
\centerline{\psfig{file=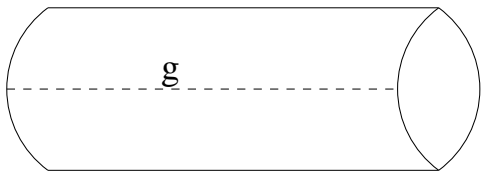,width=2.5in}}
\caption{\label{c2} An annulus diagram in an orbifold }
\end{figure}

In the case of an orbifold, the relevant annulus diagram is as
shown in figure~\ref{c2}, and has a branch cut running between
the boundary states.  We can interpret this as either an open
string propagating in a loop, coming back to itself up to the
action of some element $g$, or alternately as a closed string
in the $g$ twisted sector, propagating between two boundary states.

The analysis of {\it e.g.} \cite{bcr} also applies to the case of
noneffective orbifolds, and allows us to give nontrivial $G$-actions
to Chan-Paton factors even if $G$ acts trivially on the base,
so long as we are careful to count all boundary states.
For example, in $[X/{\bf Z}_n]$ where the ${\bf Z}_n$ acts trivially
on $X$, then even though $G$ acts trivially on $X$,
we still must distinguish boundary states in each
of $| {\bf Z}_n | = n$ twisted
sectors, even though many of these boundary states appear otherwise
identical, just as the closed string massless spectrum is $n$ copies
of the untwisted sector.

Mathematically, the combination of a trivial action on the underlying space
and a nontrivial action on the Chan-Paton factors means that B-branes
in such orbifolds are twisted sheaves on the underlying space.
In fact, there is a general statement that sheaves on gerbes are the
same as twisted sheaves on the underlying space.  These matters will
be discussed in detail in \cite{tonyme}.

\section{Mirror symmetry for completely trivial group actions}  \label{mirrors}

In section~\ref{trivgcft}, we argued that the CFT of a an
orbifold of a sigma model on $X$
by a completely trivially-acting group $G$,
{\it i.e.} a trivial
$G$-gerbe over a Calabi-Yau $X$, decomposes as a tensor product
\begin{displaymath}
{\cal C}_X \otimes {\cal C}_G
\end{displaymath}
where ${\cal C}_X$ is the CFT associated to a sigma model on $X$,
and ${\cal C}_G$ is the $G$-orbifold of a single point, $[\mbox{point}/G]$.
                                                                                
Given that result, we can immediately read off how mirror symmetry
must work for trivial gerbes over spaces.
If $X$ and $Y$ are a mirror pair of Calabi-Yau manifolds,
then by definition of mirror symmetry, ${\cal C}_X \cong {\cal C}_Y$,
hence
\begin{displaymath}
{\cal C}_X \otimes {\cal C}_G \: \cong \: {\cal C}_Y \otimes {\cal C}_G
\end{displaymath}
so we have that the trivial $G$-gerbe on $X$ is mirror to the trivial
$G$-gerbe on $Y$.

It is very easy to check that this prediction is compatible with
massless spectra.  Recall that if $X$ and $Y$ have
complex dimension $n$, then their Hodge numbers satisfy
\begin{displaymath}
h^{i,j}(X) \: = \: h^{n-i,j}(Y)
\end{displaymath}
Now, the massless spectrum of
the trivial gerbe $[X/G]$ ({\it i.e.} $G$ acts trivially on $X$)
is just copies of the massless spectrum of $X$, one copy for each
conjugacy class of $G$, with $U(1)_R$ charges and conformal weights
unchanged, hence we immediately have the trivial result
that
\begin{displaymath}
h^{i,j}([X/G]) \: = \: h^{n-i,j}([Y/G])
\end{displaymath}
confirming our claim above.

We shall discuss mirror symmetry for noneffective quotients and stacks
much more extensively in \cite{glsm}.

\section{Conclusions}

In this paper we have discussed some basic features of noneffective
orbifolds, {\it i.e.} orbifolds in which nontrivial elements of the
orbifold group act trivially.  We have seen that the resulting physical
theories are very different from orbifolds by effectively-acting groups.
We have discussed their consistency in a variety of examples,
studied closed string massless spectrum computations in detail, 
discussed D-branes in such orbifolds, and looked at some of the special
properties of orbifolds in which all elements of the group act trivially.

An important issue in understanding such gauged sigma models is
the interpretation of the moduli fields.  There are typically
more (unobstructed) moduli fields than there are geometric moduli; 
how are the rest
interpreted?
Understanding the resolution of this puzzle has led us to a class of CFT's
with a novel description, in terms of fields valued in roots of unity,
an algebraic description of twist fields associated to trivially-acting
group elements.

We will return to these issues in \cite{tonyme,glsm}, where we will
describe a more complete classification of universality classes of
worldsheet RG flow of gauged sigma models.  There, we will find general
results for things ranging from massless spectra of IR fixed points 
to mirror symmetry, and will describe some of the new physics that
arises in such considerations, such as examples of multiple
distinct nonperturbative completions
of perturbative two-dimensional gauge theories, and more independent
derivations  
of fields valued in roots of unity.

All of this has a mathematical interpretation in terms of stacks,
as we have begun to outline in this paper.

\section{Acknowledgements} 

We would like to thank A.~Adams, J.~Distler, S.~Katz, A.~Knutson,
J.~McGreevy, and
R.~Plesser for useful conversations.  We would also like to thank the
Aspen Center for Physics for hospitality while this work was being
done and the UPenn Math-Physics group for the excellent 
conditions for collaboration it provided during several stages of this
work.  T.P. was partially supported by NSF grants  DMS
0403884 and FRG 0139799.

%\bibliographystyle{my-h-elsevier}
%\bibliography{eric}

\end{document}